\documentclass[11pt,reqno]{amsart}
\usepackage{amssymb}

\newcommand{\metric}{\mathrm{g}}

\begin{document}

\title[]{A scalar-tensor cosmological model with dynamical light velocity}%

\author{M.~A.~Clayton}%
\address[1]{Department of Physics \\
Acadia University \\
Wolfville, Nova Scotia \\
B$0$P $1$X$0$, Canada}
\email{michael.clayton@acadiau.ca}
\author{J.~W.~Moffat}%
\address[2]{Department of Physics \\
University of Toronto \\
Toronto, Ontario \\
M$5$S $1$A$7$, Canada.} \email{moffat@medb.physics.utoronto.ca}

\date{\today}




\begin{abstract}
The dynamical consequences of a bimetric scalar-tensor theory of
gravity with a dynamical light speed are investigated in a
cosmological setting.  The model consists of a minimally-coupled
self-gravitating scalar field coupled to ordinary matter fields in
the standard way through the metric:
$\metric_{\mu\nu}+B\partial_\mu\phi\partial_\nu\phi$.  We show
that in a universe with matter that has a radiation-dominated
equation of state, the model allows solutions with a de~Sitter
phase that provides sufficient inflation to solve the horizon and
flatness problems. This behaviour is achieved without the addition
of a potential for the scalar field, and is shown to be largely
independent of its introduction.  We therefore have a model that
is fundamentally different than the potential-dominated,
slowly-rolling scalar field of the standard models inflationary
cosmology. The speed of gravitational wave propagation is predicted to be
significantly different from the speed of matter waves and photon propagation
in the early universe.
\end{abstract}

\maketitle

\section{Introduction}

Recently we have introduced a class of models that could loosely
be described as a minimal introduction of dynamically evolving
propagation speeds for different fields. These
models~\cite{Clayton+Moffat:1999,Clayton+Moffat:1999a,Clayton+Moffat:2000}
use a prior-geometric combination of fields to form a metric that
is universally coupled to matter fields, thereby not introducing
violations of the weak and Einstein equivalence principles.  They
are essentially a dynamical realization of the earlier
work~\cite{Moffat:1993a,Moffat:1993b,Moffat:1998} by one of the
present authors on the possibility of a phase transition in the
speed of light in the very early universe solving the horizon and
flatness problems of cosmology.

The model considered herein is that introduced
in~\cite{Clayton+Moffat:1999a}, and uses a scalar field $\phi$
that is minimally coupled to a gravitational field described by
the metric $\metric_{\mu\nu}$, to construct the matter metric:
\begin{equation}\label{eq:bimetric}
 \hat{\metric}_{\mu\nu}=\metric_{\mu\nu}+B\partial_\mu\phi\partial_\nu\phi.
\end{equation}
It is this metric that is used to construct the matter action and
can be said to be the geometry on which matter fields propagate.
Throughout we will refer to $\metric_{\mu\nu}$ as the
gravitational metric, $\hat{\metric}_{\mu\nu}$ as the matter
metric, and since $\phi$ is being used to introduce a bimetric
structure through~\eqref{eq:bimetric}, we shall refer to $\phi$ as
the biscalar field in order to distinguish it from other scalar
fields that may appear in the matter model.  It is the combination
of the gravitational metric and the biscalar field that we
consider as being the gravitational fields of our theory.

This type of model has attracted some interest in the
field~\cite{Brandenberger+Magueijo:1999,Liberati+:2000,Bassett+:2000},
and similar models that introduce the prior-geometric structure in
a different way have also
appeared~\cite{Drummond:1999,Drummond:2000}.  Since a varying
light velocity (in $3+1$ spacetime) is a fairly generic feature of
higher-dimensional
models~\cite{Kiritis:1999,Alexander:1999,Csaki+Erlich+Grojean:2000},
they are also of interest as effective or toy models that may
result from such a theory. There is also the more
phenomenologically-based `varying constants'
theories~\cite{Albrecht+Maguiejo:1999,Barrow:1999,Barrow+Maguiejo:1999,Maguiejo:2001},
and investigation of the quantum implications of such models in
the very early
universe~\cite{Harko+Mak:1999,Harko+Mak:1999a,Harko+:2000}.  Lest
one think that these models are purely of theoretical interest, an
analysis of quasar spectra~\cite{Webb+:1998,Webb+:2000} and of the
cosmic microwave background~\cite{Avelino+:2000} have both
indicated the possibility of time-variation of the fine structure
constant, and placed constraints on the variation of the speed of
light since last
scattering~\cite{Avelino+Martins+Rocha:2000,Avelino+Martins:1999}.

In this work we will explore the possibility that our model
contains cosmological solutions with sufficient inflation to solve
the horizon and flatness problems. We will begin in
Section~\ref{sect:model intro} by reviewing the model introduced
in~\cite{Clayton+Moffat:1999a} and its reduction to a homogeneous
and isotropic spacetime. As we shall see, the resulting field
equations are quite different than those of standard Einstein
gravity plus matter or scalar-tensor models, and in
Section~\ref{sect:biscalar only} we present an (implicitly
defined) solution for vanishing matter. We find that prior to a
time $t_{\mathrm{pt}}$, (massless) matter fields will continue to
propagate with a velocity $c$ (we will work in a comoving frame of
the matter metric $\hat{\metric}_{\mu\nu}$), whereas the speed at
which gravitational and biscalar field disturbances propagate,
which is determined by $\metric_{\mu\nu}$, will become vanishingly
small. This `splitting' of the two light cones causes the universe to 
experience exponential inflation, without requiring the existence
of a self-interaction potential for the biscalar field. At
$t_{\mathrm{pt}}$ the restoration of Lorentz symmetry occurs, and
the light cones of all fields coincide. Assuming that
$t_{\mathrm{reh}}\ge t_{\mathrm{pt}}$ (the former is the reheating
time), we show that we can have sufficient inflation to solve the
horizon and flatness problems.

In Section~\ref{sect:pcosm} we re-introduce matter with a
radiation equation of state, and find an approximate solution
that, in addition to the symmetry restoration time
$t_{\mathrm{pt}}$, has an earlier period ($t<t_{\mathrm{t}}$)
during which the light cones of the gravitational and biscalar
field are also `split'.

One fundamental issue that we will encounter is how to define
``the Planck scale''.  Combinations of quantities such as, for
example, $\ell_{\mathrm{P}}=\sqrt{G\hbar/c^3}$, depend on the
speed of light $c$, gravitational coupling $G$ and Planck's
constant $\hbar$.  Since in our model we have different
propagation speeds for different fields (which are also
dynamically varying) and we can expect that the effective
gravitational coupling to matter also changes with time, we should
expect that the scale at which quantum effects become important
will be different for different fields.  It is also normally
assumed that $\hbar$ is a fixed constant, but clearly if we are
allowing fundamental constants to vary with time, then it is not
unreasonable to consider it to be represented by a field:
$\hbar(t,\vec{x})$. Although we will not completely resolve these
issues, we will motivate the choice of the parameter $B$ appearing
in~\eqref{eq:bimetric} as
\begin{equation}
B\approx\frac{1}{32\pi}\ell_{\mathrm{P}}^2.
\end{equation}

In Section~\ref{sect:pcosm} we show that including matter with a
radiation equation of state will not alter the de~Sitter behaviour
significantly, but it will introduce another time scale
$t_{\mathrm{t}}$ at which the light cone of the biscalar field
deviates from that of the gravitational field $\metric_{\mu\nu}$.
We therefore see that in general we will have two time scales in
the model, each associated with the bifurcation of light cones
associated with different fields.
In Section~\ref{sect:conclusion}, we end with concluding remarks.

\section{The model}
\label{sect:model intro}

The model that we introduced in~\cite{Clayton+Moffat:1999a}
consisted in a self-gravitating scalar field coupled to matter
through the matter metric~\eqref{eq:bimetric}, with action
\begin{equation}
S=S_{\rm grav}+S_{\phi}+\hat{S}_{\rm M},
\end{equation}
where
\begin{equation}
S_{\rm grav}=-\frac{1}{\kappa}\int d\mu (R[\metric]+2\Lambda),
\end{equation}
$\kappa=16\pi G/c^4$, $\Lambda$ is the cosmological constant, and
we employ a metric with signature $(+,-,-,-)$.  We will write, for
example, $d\mu=\sqrt{-\metric}\;d^4x$ and $\mu=\sqrt{-\metric}$
for the metric density related to the gravitational metric
$\metric_{\mu\nu}$, and similar definitions of $d\hat{\mu}$ and
$\hat{\mu}$ in terms of the matter metric
$\hat{\metric}_{\mu\nu}$.  The minimally-coupled scalar field
action is\footnote{Note that our earlier
publication~\cite{Clayton+Moffat:1999a} contained a typo in
equation~($5$); the rest of the letter is correct as published.}:
\begin{equation} S_{\rm \phi}=\frac{1}{\kappa}\int d\mu\,
\Bigl[\frac{1}{2}\metric^{\mu\nu}\partial_\mu\phi\partial_\nu\phi-V(\phi)\Bigr],
\end{equation} where the scalar field $\phi$ has been chosen to be dimensionless. The
energy-momentum tensor for the scalar field that we will use is
given by
\begin{equation}
T^{\mu\nu}_\phi=\frac{1}{\kappa}\Bigl[
\metric^{\mu\alpha}\metric^{\nu\beta}\partial_\alpha\phi\partial_\beta\phi
-\tfrac{1}{2}\metric^{\mu\nu}\metric^{\alpha\beta}\partial_\alpha\phi\partial_\beta\phi
+\metric^{\mu\nu}V(\phi) \Bigr],
\end{equation}
and is the variation of the scalar field action with respect to
the gravitational metric: $\delta S_{\phi}/\delta \metric_{\mu\nu}
 =-\frac{1}{2}\mu T_\phi^{\mu\nu}$.

Instead of constructing the matter action $\hat{S}_{\mathrm{M}}$
using the metric $\metric_{\mu\nu}$, we use the
combination~\eqref{eq:bimetric} resulting in the identification of
$\hat{\metric}_{\mu\nu}$ as the physical metric that provides the
arena on which matter fields interact. That is, the matter action
$\hat{S}_{\mathrm{M}}[\psi^I] =
\hat{S}_{\mathrm{M}}[\hat{\metric},\psi^I]$, where $\psi^I$
represents all the matter fields in spacetime, is one of the
standard forms, and therefore the energy-momentum tensor derived
from it by
\begin{equation}\label{eq:matterEM}
\frac{\delta S_{\mathrm{M}}}{\delta \hat{\metric}_{\mu\nu}}
 =-\frac{1}{2}\hat{\mu}\hat{T}^{\mu\nu},
\end{equation}
satisfies the conservation laws
\begin{equation}\label{eq:matterconservation}
\hat{\nabla}_\nu\Bigl[\hat{\mu}\hat{T}^{\mu\nu}\Bigr]=0,
\end{equation}
as a consequence of the matter field equations
only~\cite{Clayton+Moffat:1999a}.  It is the matter covariant
derivative $\hat{\nabla}_\mu$ that appears here, which is the
metric compatible covariant derivative determined by the matter
metric: $\hat{\nabla}_\alpha\hat{\metric}_{\mu\nu}=0$.

As described in~\cite{Clayton+Moffat:1999a}, the gravitational
field equations for this model can be written as
\begin{equation}\label{eq:Einsteins eqns}
 G^{\mu\nu}=\Lambda\metric^{\mu\nu}
 +\frac{\kappa}{2}T^{\mu\nu}_\phi
 +\frac{\kappa}{2}\frac{\hat{\mu}}{\mu}\hat{T}^{\mu\nu},
\end{equation}
and that for the scalar field (written here in terms of matter
covariant derivatives) as:
\begin{equation}\label{eq:scalar FEQ}
\bar{\metric}^{\mu\nu}\hat{\nabla}_\mu\hat{\nabla}_\nu\phi+KV^\prime
[\phi]=0.
\end{equation}
In the latter, we have defined the biscalar field metric
\begin{equation}\label{eq:scalar metric}
 \bar{\metric}^{\mu\nu} = \hat{\metric}^{\mu\nu}
 +\frac{B}{K}\hat{\nabla}^\mu\phi\hat{\nabla}^\nu\phi
 -\kappa\frac{\hat{\mu}}{\mu}BK\hat{T}^{\mu\nu},
\end{equation}
and
\begin{equation}\label{eq:K definition}
 K=1-B\hat{\metric}^{\mu\nu}\partial_\mu\phi\partial_\nu\phi.
\end{equation}
In this work we will assume a perfect fluid form for the matter
fields:
\begin{equation}
 \hat{T}^{\alpha\beta}=
 \Bigl(\rho+\frac{p}{c^2}\Bigr)\hat{u}^\alpha\hat{u}^\beta
 -p\hat{\metric}^{\alpha\beta},
\end{equation}
with $\hat{\metric}_{\mu\nu}\hat{u}^\mu\hat{u}^\nu=c^2$.

We will now specialize to a cosmological setting, imposing
homogeneity and isotropy on spacetime and writing the matter
metric in comoving form as:
\begin{equation}
 \hat{\metric}_{\mu\nu}=\mathrm{diag}(c^2,-R^2(t)\gamma_{ij}),
\end{equation}
with coordinates $(t, x^i)$ and $3$-metric $\gamma_{ij}$ on the
spatial slices of constant time. The matter stress-energy tensor
(using $\hat{u}^0=c$):
\begin{equation}
 \hat{T}^{00}=\rho,\quad
 \hat{T}^{ij}=\frac{p}{R^2}\gamma^{ij},
\end{equation}
then leads to the conservation laws (an overdot indicates a
derivative with respect to the time variable $t$, and
$H=\dot{R}/R$ is the Hubble function)
\begin{equation}\label{eq:cosm cons}
\dot{\rho}+3H\Bigl(\rho+\frac{p}{c^2}\Bigr)=0.
\end{equation}
Since we are interested in the very early universe, we will assume
a radiation equation of state:
\begin{equation}\label{eq:equation of state}
p=\frac{1}{3}c^2\rho,
\end{equation}
which leads to $\rho\propto 1/R^4$.

It is useful at this point to introduce the following quantities
derived from the constant $B$ which will appear throughout this
work:
\begin{equation}\label{eq:definitions}
H_B^2=\frac{c^2}{12B},\quad
\rho_B=\frac{1}{2\kappa c^2 B},
\end{equation}
the latter comes from $H_B^2=\frac{1}{6}\kappa c^4\rho_B$.  From
the definition~\eqref{eq:K definition} we have
\begin{equation}\label{eq:K cosm}
 K=1-\frac{\dot{\phi}^2}{12H_B^2},
\end{equation}
and using the relation~\eqref{eq:bimetric} the gravitational
metric is found to be
\begin{equation}
 {\metric}_{\mu\nu}=\mathrm{diag}(K c^2,-R^2\gamma_{ij}),
\end{equation}
and so $\mu=\sqrt{K}\hat{\mu}$.  From the
definition~\eqref{eq:scalar metric} we find:
\begin{equation}\label{eq:g bar}
 \bar{\metric}^{\mu\nu}=\mathrm{diag}\Bigl(\frac{1}{K c^2}{I}_\rho,
 -\frac{1}{R^2}\gamma^{ij}{I}_p\Bigr),
\end{equation}
where we have defined
\begin{equation}
 {I}_\rho = 1-K^{3/2}\frac{\rho}{2\rho_B},\quad
 {I}_p = 1 + \sqrt{K}\frac{\rho}{6\rho_B}.
\end{equation}
In this work, we will find that neither $I_\rho$ nor $I_p$ vanish
in the solutions of interest, so the metric~\eqref{eq:scalar
metric} is degenerate only when $R=0$ or $K=0$.

The Friedmann equation ($R_{00}+Kc^2\gamma^{ij}R_{ij}$) is given
by
\begin{subequations}\label{eq:grav eqns}
\begin{equation}\label{eq:Fried eqn}
 H^2 +\frac{kc^2K}{R^2} =
 \frac{1}{3}Kc^2\Lambda
 +H_B^2(1-K)
 +2H_B^2 K V_B
 +\frac{1}{6}\kappa c^4 K^{3/2}\rho.
\end{equation}
The remaining equation from~\eqref{eq:Einsteins eqns} is
\begin{equation}\label{eq:ddot eqn}
 2\dot{H}
 +3H^2
 -H\frac{\dot{K}}{K}
 +\frac{kc^2K}{R^2}
 =c^2\Lambda K
 -3H_B^2(1-K)
 +6H_B^2 K  V_B
 -\frac{1}{6}\kappa c^4\sqrt{K}\rho,
\end{equation}
\end{subequations}
which can be shown to be redundant given the scalar field equation
(below), the matter conservation laws~\eqref{eq:cosm cons} and the
Friedmann equation.  The scalar field equation~\eqref{eq:scalar
FEQ} in this gauge is
\begin{subequations}\label{eq:biscalar equations}
\begin{equation}\label{eq:phi dot}
 {I}_\rho\ddot{\phi}
 +3KH{I}_p\dot{\phi}
 +c^2 K^2 V^\prime
 =0,
\end{equation}
which, using~\eqref{eq:K cosm} can also be written as
\begin{equation}\label{eq:K dot}
 {I}_\rho\dot{K}
 -6K(1-K)H{I}_p
 -2 K^2 \dot{V}_B
 =0.
\end{equation}
\end{subequations}
We have introduced the dimensionless potential:
$V_B=BV$ and $\dot{V}_B=V^\prime_B\dot{\phi}$.

These combined equations suggest the following definitions
motivated by the varying constants
theory~\cite{Albrecht+Maguiejo:1999,Barrow:1999}:
\begin{equation}\label{eq:vc}
 c_{\mathrm{vc}}(t)=\sqrt{K}c,\quad
 G_{\mathrm{vc}}(t)=K^{3/2}G,
\end{equation}
in which case the Friedmann equation becomes
\begin{equation}\label{eq:modified Fried}
 H^2 +\frac{kc^2_{\mathrm{vc}}}{R^2} =
 \frac{1}{3}c^2_{\mathrm{vc}}\Lambda
 +\frac{1}{12}\dot{\phi}^2
 +\frac{1}{6}c^2 K V
 +\frac{8\pi}{3}G_{\mathrm{vc}}\rho.
\end{equation}
Aside from the biscalar potential term (which for simple
polynomial potentials can be re-written as a potential with time
varying mass and coupling-constants) we have a \textit{form} that
is identical to the Friedmann equation for matter and a scalar
field minimally coupled to Einstein gravity.  This suggests the
definition of a Planck scale that is no longer a constant
quantity, but varies as, for example:
\begin{equation}\label{eq:vc Planck density}
 \rho_{\mathrm{P,vc}}(t)
 =\frac{c^5_{\mathrm{vc}}(t)}{\hbar G^2_{\mathrm{vc}}(t)}
 =\frac{1}{\sqrt{K}}\rho_{\mathrm{P}},
\end{equation}
where $\rho_{\mathrm{P}}= c^5/(\hbar G^2)$.

We should note that in~\cite{Avelino+Martins:1999} it is stated
that varying constants theories require a violation of the strong
energy condition in order to solve the horizon and flatness
problems.  Since we will be setting the biscalar field potential
to zero, one might ask how it is possible to claim to solve these
problems with matter that satisfies a radiation equation of state?
The answer to this is that the Friedmann equation~\eqref{eq:Fried
eqn} does in fact display coupling to sources that appear to
violate the strong energy condition.  Note though that to a
material observer, matter fields \textit{will} obey the strong
energy condition but couple to the gravitational field in a way
that does not.  Thus it is the reaction of the gravitational field
that sees such violations, which is precisely what is necessary
for inflation.

\section{empty space, biscalar field model}
\label{sect:biscalar only}

The simplest possible scenario is one in which there is no matter
present, and the biscalar field is solely responsible for the
dynamics of the very early universe.  As we will see in the
following section, the presence of matter will have a nontrivial
effect at earliest times, but not in a way that will affect
the scenario significantly.

From~\eqref{eq:biscalar equations} we see that both $K=0$ and
$K=1$ are solutions, the first corresponding to an exact de~Sitter
solution and the latter to a model in which the biscalar field
plays no active role.  It is not hard to show that the $K=0$
solutions are unstable, with deviations $K\propto R^6$, and so we
end up with a strong relationship between how small $K$ is at the
beginning of the universe and the amount of inflation achieved.
This fixed point corresponds to degenerate $\metric_{\mu\nu}$ and
$\bar{\metric}_{\mu\nu}$, and so physical solutions will have
$K>0$ except possibly at a point, which could be considered a
physical singularity.  What we are essentially going to find here
are solutions that interpolate between these two limits, beginning
near $K=0$ and making a transition (at roughly $t_{\mathrm{pt}}$)
to a minimally-coupled scalar field model.

Setting $\rho=0=p$ and $V_B=0$ in~\eqref{eq:K dot}, we find the
solution
\begin{equation}\label{eq:K bimetric}
K=\Bigl(1+2\frac{R^6_{\mathrm{pt}}}{R^6}\Bigr)^{-1},
\end{equation}
where we have defined a scale $R_{\mathrm{pt}}$ to correspond to
the end of the inflationary phase, at which point we will have
$K_{\mathrm{pt}}=1/3$.  This is justified by using~\eqref{eq:grav
eqns} and~\eqref{eq:K dot} to derive
\begin{equation}
\ddot{R}=RH_B^2(1-K)(1-3K),
\end{equation}
which demonstrates that we have inflation ($\ddot{R}>0$) when
$K<1/3$.  The subscript ``$\mathrm{pt}$'' indicates that this is
the scale at which the phase transition between an earlier `broken
phase' where the propagation speeds for matter and gravitation are
very different, and a later `restored phase' where these speeds
become approximately equal.

Using~\eqref{eq:K bimetric} in the Friedmann
equation~\eqref{eq:Fried eqn}, we find the implicit solution
\begin{equation}\label{eq:implicit}
 \sqrt{2(2+y)}-\ln\Biggl(\frac{\sqrt{2(2+y)}+2}{\sqrt{2(2+y)}-2}\Biggr)
 -y_c=6H_B(t-t_{\mathrm{pt}}),
\end{equation}
where
\begin{equation}
y=\Bigl(\frac{R}{R_{\mathrm{pt}}}\Bigr)^6,\quad
y_c=\sqrt{6}-\ln\Bigl(\frac{\sqrt{6}+2}{\sqrt{6}-2}\Bigr)\approx
1.57,
\end{equation}
with $y_c$ resulting from fixing the constant of integration so
that $R=R_{\mathrm{pt}}$ at $t=t_{\mathrm{pt}}$.

For $R\ll R_{\mathrm{pt}}$ we can expand~\eqref{eq:implicit} in
$y$ to find
\begin{equation}
R\approx R_{\mathrm{pt}}\exp\bigl[\tfrac{1}{6}(y_c+\ln
8-2)\bigr]e^{H_B(t-t_{\mathrm{pt}})}
\approx R_{\mathrm{pt}}e^{H_B(t-t_{\mathrm{pt}})},
\end{equation}
and since $\exp\bigl[\tfrac{1}{6}(y_c+\ln 8-2)\bigr]\approx 1.27$
we use this approximation right up to $t_{\mathrm{pt}}$.  In the
opposite limit ($R\gg R_{\mathrm{pt}}$), we find that
\begin{equation}
R\approx
R_{\mathrm{pt}}\Bigl[\frac{z_c}{\sqrt{2}}+3\sqrt{2}H_B(t-t_{\mathrm{pt}})\Bigr]^{1/3}
\approx
R_{\mathrm{pt}}\Bigl[1+3\sqrt{2}H_B(t-t_{\mathrm{pt}})\Bigr]^{1/3},
\end{equation}
where in this case we are making only a slightly larger error
since $y_c/\sqrt{2}\approx 0.111$.

In the very early universe ($t\ll t_{\mathrm{pt}}$) the Friedmann
equation has a source that is effectively a constant energy
density $\rho_B$, so it is reasonable to assume that
$\rho_B\approx\rho_P=c^5/(\hbar G^2)$.  At $t_{\mathrm{pt}}$ the
effective energy density has fallen to $2/3$ of this value (since
$K_{\mathrm{pt}}=1/3$) and at some scale following inflation we
expect reheating to take place. It is likely that in order to have
reheating we will need to consider a non-vanishing biscalar field
potential, which for concreteness we will assume is a simple
quadratic form $V(\phi)\approx 16\pi l^{-2}\phi^2$ where $l$ is a
length scale characterizing the mass of the biscalar field. The
dimensionless potential is then $V_B=16\pi
Bl^{-2}\phi^2$ with $32\pi Bl^{-2}\approx(\ell_{\mathrm{P}}/l)^2$
a dimensionless parameter.

Reheating will begin when the biscalar field begins to oscillate
about $\phi=0$, which is characterized by the potential dominating
the Hubble damping.  It is possible that this could happen prior
to $t_{\mathrm{pt}}$, in which case the analysis becomes quite
complicated, since one would have to consider nonlinear
contributions to~\eqref{eq:biscalar equations}. If it happens at
or after $t_{\mathrm{pt}}$, then we can approximate the scalar
field equation by
\begin{equation}
 \ddot{\phi}
 +3H\dot{\phi}
 +32\pi(c/l)^2 \phi
 \approx 0,
\end{equation}
and find that reheating begins at a scale $R_{\mathrm{reh}}$ where
$H^2_{\mathrm{reh}}\approx 32\pi\times4c^2/(9l^2)$.  From this we
see that reheating begins at or after $t_{\mathrm{pt}}$ if $l^2\ge
8(32\pi B)\approx 8\ell_{\mathrm{P}}^2$.  We will leave the
details of reheating to future work, and assume for simplicity
that reheating takes place at $t_{\mathrm{reh}}=t_{\mathrm{pt}}$
and is instantaneous. This means that $\rho_{\mathrm{reh}}$, the
radiation energy density produced at the end of reheating, should
satisfy $H_B^2(1-K_{\mathrm{pt}})=\frac{1}{6}\kappa c^4
\rho_{\mathrm{reh}}$, which gives
$\rho_{\mathrm{reh}}=\frac{2}{3}\rho_B$.  With this assumption, we
would have the reheating temperature not too different from the
Planck temperature: $T_{\mathrm{pt}}\approx
T_{\mathrm{reh}}\approx T_{\mathrm{P}}$.

The horizon problem is solved, if the size of the cosmological
horizon
\begin{equation}
r_c = \frac{cR_0}{RH},
\end{equation}
at some time in the history of the universe is much greater than
its value at present~\cite{Coles+Luccin:1995}, that is: $\max_t
r_c(t)\gg r_c(t_0)$.  During inflation ($t<t_{\mathrm{pt}}$) the
cosmological horizon is decreasing, and it will have a maximum
value at earlier times.  As it stands our purely classical model
can be traced back to $t=-\infty$ and in that limit
$r_c\rightarrow \infty$.  More realistically, the model becomes a
good approximation to a more fundamental quantum model after a
time we will call $t_{\mathrm{qg}}$.  We then must have
$r_c(t_{\mathrm{qg}})\gg r_c(t_0)$, which yields the condition
$R_0H_0\gg R_{\mathrm{qg}}H_{\mathrm{qg}}$, or
\begin{equation}
\frac{R_0H_0}{R_{\mathrm{pt}}H_{\mathrm{pt}}} \gg
\frac{R_{\mathrm{qg}}H_{\mathrm{qg}}}{R_{\mathrm{pt}}H_{\mathrm{pt}}}.
\end{equation}
Assuming that the universe undergoes $\mathcal{N}$ e-folds in this
time, we have $R_{\mathrm{pt}}/R_{\mathrm{qg}}=\exp(\mathcal{N})$,
$K_{\mathrm{qg}}\approx \frac{1}{2}e^{-6\mathcal{N}}$ and
$H_{\mathrm{pt}}/H_{\mathrm{qg}}\approx \sqrt{2/3}$. Using
standard physics following $t_{\mathrm{pt}}\approx
t_{\mathrm{reh}}$, this constraint then becomes
\begin{equation}
\mathcal{N}\gg
\ln\Bigl(\frac{T_{\mathrm{pt}}}{T_{\mathrm{P}}}\sqrt{10^{60}z_{\mathrm{eq}}}\Bigr)
\in (32,60).
\end{equation}
The redshift at matter-radiation equality is
$z_{\mathrm{eq}}\approx 4.3\times 10^4\Omega h^2 \approx 4\times
10^4$, and $1< T_{\mathrm{sr}}/T_{\mathrm{P}}< 10^{-5}$.  Thus we
have the usual statement that $\mathcal{N}\approx 60$ will solve
the horizon problem.

The demonstration that this also solves the flatness problem
proceeds in a manner similar to that described
in~\cite{Clayton+Moffat:2000}. Identifying the physically
meaningful measure of the flatness of a spacelike hypersurface as
the $3$-curvature measured by a material observer, we have (see
also~\cite{Hu+Turner+Weinberg:1994})
\begin{equation}
\lvert\Omega-1\rvert=\frac{kc^2}{R^2H^2}.
\end{equation}
We see that this increases without bound as $R\rightarrow 0$, but
relating the flatness at $t_{\mathrm{qg}}$ and $t_{\mathrm{pt}}$
we find
\begin{equation}
\frac{\lvert\Omega-1\rvert_{\mathrm{qg}}}
{\lvert\Omega-1\rvert_{\mathrm{pt}}}
 =\frac{R_{\mathrm{pt}}^2H_{\mathrm{pt}}^2}{R_{\mathrm{qg}}^2H_{\mathrm{qg}}^2}
 \approx \frac{2}{3}e^{2\mathcal{N}},
\end{equation}
and we see that solving the horizon problem implies that we have
also solved the flatness problem, as expected.

Using the definition~\eqref{eq:K cosm} and the
solution~\eqref{eq:K bimetric} we can determine how `far' the
scalar field travels between two scales $R_i$ and $R_f$:
\begin{equation}
\lvert\Delta\phi\rvert =2\sqrt{3}\ln(R_f/R_i),
\end{equation}
and so between $t_{\mathrm{qg}}$ and $t_{\mathrm{pt}}$ we have
$\lvert\Delta\phi\rvert = 2\sqrt{3}\mathcal{N}\approx 200$.  Note
though, that even if the scalar field has a relatively large value
at $t_{\mathrm{qg}}$, the effect of the potential is essentially
removed by the factor involving $K$ that multiplies it.  Thus the
effect of the potential is delayed until after $t_{\mathrm{pt}}$,
and the magnitude of the scalar field will have an effect on
reheating rather than the dynamics of inflation. We therefore see
that in this model the biscalar field can in fact `roll' quite
far, but since the potential (and therefore the value of $\phi$)
has essentially no effect on the de~Sitter behaviour, we do not
need an extraordinarily flat potential to achieve sufficient
inflation.  (Because it is essentially the nonstandard kinetic
terms for the biscalar field that are driving the de~Sitter phase,
our model shares some similarities with pre-big bang string
cosmology~\cite{Veneziano:1998}.)  This makes us believe that this
model may have something rather significant to say about the
cosmological constant problem, since vacuum energy in the very
early universe should be irrelevant to inflation in this model.

On the other hand, we do require that at $t_{\mathrm{qg}}$ the
value of $K$ is extremely small: $K_{\mathrm{qg}}\approx
\exp(-6\mathcal{N})$, which could be viewed as a form of
fine-tuning. There are a couple of ways of looking at this
problem.  As an initial conditions problem, we really only need to
argue that in some more fundamental theory we expect that the
universe should enter this classical regime close to a vacuum
where the light cones are quite different.  This is probably not
unreasonable, for we have seen in the literature
dimensionally-reduced models with light cone variability similar
to that of our model. Tracing our model back to this other vacuum
we are forced towards $K=0$, and so the issue becomes one of how
this more fundamental theory reduces to a model similar to ours in
the very early universe.  Alternatively, we can transform the
model back to a frame (with time coordinate $T$) in which the
gravitational metric $\metric_{\mu\nu}$ is in comoving form, using
$3H_BT\approx\exp(3H_Bt)$.  From this we find that $\partial_t\phi
\approx T\partial_T\phi$, and we see that the apparent fine-tuning
$\partial_t\phi$ is at least partially alleviated by the fact that
$T\rightarrow 0$ in the very early universe.

Although we have tentatively identified $\rho_B\approx
\rho_{\mathrm{P}}$, it is not at all clear that this is
appropriate.  The reason is that since we have fields that have a
varying propagation speed, it is perhaps more reasonable to define
the Planck scale as (assuming that $\hbar$ remains constant), for
example:
\begin{equation}\label{eq:bar rhoP}
\bar{\rho}_{\mathrm{P}}=\frac{\bar{c}^5(t)}{\hbar\bar{G}^2(t)},
\end{equation}
where $\bar{c}(t)$ and $\bar{G}(t)$ are the effective speed of
propagation and gravitational coupling of the field in question.
We would then use this (or similar definitions for other Planck
scale quantities) to argue that quantum corrections become
important at this scale.

We know from the field equation~\eqref{eq:scalar FEQ} that the
metric $\bar{\metric}_{\mu\nu}$ determines the speed of
propagation of the biscalar field, which reduces to
\begin{equation}\label{eq:biscalar speed}
\bar{c}=\frac{I_p}{I_\rho}\sqrt{K}c\approx \sqrt{K}c.
\end{equation}
Since the Friedmann equation looks identical to that of a
minimally-coupled scalar field, we assume that $\bar{G}=G$, and
from~\eqref{eq:bar rhoP} we would therefore have
$\bar{\rho}_{\mathrm{P}}=K^{5/2}\rho_{\mathrm{P}}$.  Since this
vanishes as $R \rightarrow 0$ we conclude that we must have a
scale above which quantum gravity dominates.  Assuming that at
$t_{\mathrm{qg}}$ we have $\frac{1}{6}\kappa c^4
\bar{\rho}_{\mathrm{P}}(t_{\mathrm{qg}})=H_B^2(1-K_{\mathrm{qg}})$,
we find that
\begin{equation}
\rho_B =
\frac{K^{5/2}_{\mathrm{qg}}}{1-K_{\mathrm{qg}}}\rho_{\mathrm{P}}
\approx K^{5/2}_{\mathrm{qg}}\rho_{\mathrm{P}}
 \approx 2^{-5/2}e^{-15\mathcal{N}}\rho_{\mathrm{P}},
\end{equation}
which would require $\rho_B$ to be incredibly small, if the horizon
problem is to be satisfied.  The varying constants
definition~\eqref{eq:vc Planck density} yields a result that goes
in the opposite direction:
\begin{equation}
\rho_B =
\frac{1}{\sqrt{K_{\mathrm{qg}}}(1-K_{\mathrm{qg}})}\rho_{\mathrm{P}}
\approx \frac{1}{\sqrt{K_{\mathrm{qg}}}}\rho_{\mathrm{P}}
 \approx \sqrt{2} e^{3\mathcal{N}}\rho_{\mathrm{P}},
\end{equation}
which would indicate that $\rho_B$ is greatly in excess of the
usual Planck scale.  Note that this identification would also
result in quantum corrections becoming important \textit{after}
$t_{\mathrm{qg}}$, which is not reasonable.

On the other hand, because in the gravitational frame the system
is an ordinary minimally-coupled scalar field, we would perhaps
expect that the standard definition of the Planck scale would be
appropriate--as we have assumed here.  That is not to say that
this issue is closed.  Once we include matter into the system, we
could legitimately claim that the relevant Planck scale is
different for matter fields.  We expect that a perturbation
calculation will not only provide the seeds for structure
formation, it will also illuminate when quantum corrections will
dominate the very early universe (and therefore pin down
$t_{\mathrm{qg}}$ more completely).

\section{The model with non-vanishing radiation}
\label{sect:pcosm}

In the last section we described a solution for which the matter
contributions were set to zero.  In such a limit we could have
reverted back to the field equations in the ``gravitational
frame'', and chosen coordinates such that $\metric_{\mu\nu}$ is of
comoving form.  We would then find a simple scalar field equation:
$\partial_T^2\phi+3H\partial_T\phi=0$, with solution
$\partial_T\phi\propto 1/R^3$, using which, the Friedmann equation
would yield $R\propto T^{1/3}$.  To relate this solution to that
which we found in the previous section requires making the
coordinate transformation: $dt = \sqrt{1+1/(3H_BT)^2}dT$ which
puts $\hat{\metric}_{\mu\nu}$ into comoving form.  This coordinate
transformation leaves the time coordinate essentially untouched
for $T>1/(3H_B)$, and for $T\ll 1/(3H_B)$ we see that $t\propto\ln
T$, so that whereas $T\in(0,\infty)$ we have
$t\in(-\infty,\infty)$.  If we want to retain primordial matter,
this simple solution is no longer available to us since the matter
stress-energy tensor appears in the scalar field
equation~\eqref{eq:scalar FEQ}.  We have chosen to work directly
in a comoving matter frame since the interpretation of the
resulting metric is clearer, and because the field equations with
matter are slightly more manageable.

We now wish to retain the coupling to matter, choosing a radiation
equation of state appropriate to the very early universe.
From~\eqref{eq:Fried eqn} we see that if $K\approx 0$ and
$K^{3/2}\rho\approx 0$ as $R\rightarrow 0$, then the Friedmann
equation is still dominated by the biscalar field kinetic terms,
and we have the same de~Sitter behaviour for the scale factor. In
fact what we will find is that, at least when the scalar field
potential may be neglected in the very early universe, the $K=0$
fixed point is approached as $R\rightarrow 0$ in such a way that
$\sqrt{K}\rho\rightarrow \text{constant}$.

Let us begin by parameterizing the radiation energy density $\rho$
as
\begin{equation}\label{eq:radiation}
\rho=\rho_{\mathrm{sr}}\Bigl(\frac{R_{\mathrm{sr}}}{R}\Bigr)^4,
\end{equation}
where subscript ``$\mathrm{sr}$'' indicates that the quantity is
evaluated at the scale $R_{\mathrm{sr}}$ where the biscalar and
radiation contributions to the Friedmann equation would be equal.
Defining the ratio
\begin{equation}\label{eq:sigma defn}
{\sigma} = \frac{\rho_{\mathrm{sr}}}{\rho_B},
\end{equation}
this implies that we have
\begin{equation}
(1-K_{\mathrm{sr}})\approx \sigma K_{\mathrm{sr}}^{3/2},
\end{equation}
which has the small-$\sigma$ solution:
\begin{equation}\label{eq:Ksr}
K_{\mathrm{sr}}\approx 1-2{\sigma} +
\mathcal{O}({\sigma}^2).
\end{equation}
Note that the constant $\sigma$ defined in~\eqref{eq:sigma defn},
although a very convenient parameter to use in making
approximations at later times, cannot in any way be considered as
a fundamental parameter of the model.  Since we are interested in
retaining the de~Sitter phase, it is reasonable to assume that
$t_{\mathrm{sr}}>t_{\mathrm{qg}}$, and since we expect that $\rho$
could possibly be as large as $\rho_B$ at $t_{\mathrm{qg}}$ but no
larger, then it will be redshifted significantly by inflation in
order to make $\sigma$ a very small parameter.  In
turn~\eqref{eq:Ksr} tells us that
$t_{\mathrm{sr}}>t_{\mathrm{pt}}$, so that reheating happens well
before biscalar-radiation equality.

It is convenient to define the variable
\begin{equation}\label{eq:Z def}
Z=\frac{1}{2}\sqrt{K}\Bigl(\frac{R_{\mathrm{sr}}}{R}\Bigr)^4,
\end{equation}
so that we have
\begin{equation}\label{eq:Is}
{I}_\rho= 1-{\sigma} K Z,\quad
{I}_p= 1+\frac{1}{3}{\sigma} Z.
\end{equation}
Setting the biscalar field potential to zero, we have the
following equations for $K$ and $Z$:
\begin{subequations}\label{eq:VEU}
\begin{gather}\label{eq:VEU psi II}
 I_\rho\dot{K}
 =6K(1-K)H(1+\tfrac{1}{3}{\sigma} Z),\\
\label{eq:Z dot II}
 I_\rho\dot{Z}=-H(1+3K)Z(1-{\sigma} Z).
\end{gather}
\end{subequations}
The first of these equations tells us that $K$ is a nondecreasing
function of time provided that $H\ge 0$, which is certainly the
case at present and from the form of the Friedmann
equation~\eqref{eq:Fried eqn} must also be true everywhere in the
past (recall that $K\in(0,1)$).  We therefore see that $K$ should
begin near $K\approx 0$ and increase towards $K=1$ as the universe
expands.  Since we expect that $R\gg R_{\mathrm{sr}}$ and
$K\approx 1$ at present, $Z$ should be small and decreasing
towards zero at present.  We also see from~\eqref{eq:Z dot II}
that $Z$ must be decreasing everywhere in the past, and as
$R\rightarrow 0$ we must have $Z\rightarrow 1/\sigma$.  This is
important, since it shows that near $R=0$ we must have $K\propto
R^8$ rather than the $K\propto R^6$ found in
Section~\ref{sect:biscalar only}. This guarantees that the
constant $H_B^2$ is still the only important term in the Friedmann
equation as $R\rightarrow 0$.

Since $Z$ is an increasing variable, we can write
$\partial_t=\dot{Z}\partial_Z$, and using~\eqref{eq:VEU} we find
\begin{equation}
\frac{\partial K}{\partial Z}=-\frac{6K(1-K)(1+\frac{1}{3}{\sigma}
Z)} {Z(1+3K)(1-{\sigma} Z)},
\end{equation}
which can be integrated to find ($A_{\sigma}$ is a constant of
integration):
\begin{equation}\label{eq:KZ integrated}
\frac{(1-K)^4}{K}\frac{(1-{\sigma} Z)^8}
 {Z^6}=2^6A_{\sigma}^4>0.
\end{equation}
Using the definition~\eqref{eq:Z def} this becomes
\begin{equation}\label{eq:Ky}
(1-K)\Bigl[\Bigl(\frac{R}{R_{\mathrm{sr}}}\Bigr)^4-\frac{1}{2}\sigma\sqrt{K}\Bigr]^2=
A_{\sigma} K\Bigl(\frac{R}{R_{\mathrm{sr}}}\Bigr)^2,
\end{equation}
and since at $R=R_{\mathrm{sr}}$ we have $y=1$, we must have
\begin{equation}
(1-K_{\mathrm{sr}})(1-\tfrac{1}{2}{\sigma}\sqrt{K_{\mathrm{sr}}})^2=
A_{\sigma} K_{\mathrm{sr}},
\end{equation}
and combining this with~\eqref{eq:Ksr} we obtain
\begin{equation}
 A_{\sigma}
 =2{\sigma}\sqrt{K_{\mathrm{sr}}}(1-\tfrac{1}{2}{\sigma}\sqrt{K_{\mathrm{sr}}})^2
 \approx 2{\sigma} +\mathcal{O}({\sigma}^2).
\end{equation}

We shall now seek an asymptotic (in ${\sigma}$) solution
to~\eqref{eq:Ky}, and, re-summing the first term, we find
\begin{subequations}\label{eq:K approx}
\begin{equation}\label{eq:K long}
K\approx
\Bigl[1+2\sigma\Bigl(\frac{R_{\mathrm{sr}}}{R}\Bigr)^6\Bigr]^{-1},
\end{equation}
which is identical to~\eqref{eq:K bimetric} up to rescaling.
Taking $K_{\mathrm{pt}}=1/3$ as before, this tells us that
$R_{\mathrm{sr}}/R_{\mathrm{pt}}=\sigma^{1/6}$.  Note that this is
not a uniform expansion (the higher-order corrections diverge as
$R\rightarrow 0$), and we perform a series expansion in
$R/R_{\mathrm{pt}}$ to find (once again re-summing the leading
contribution):
\begin{equation}\label{eq:K small}
K\approx \Bigl(\frac{R}{R_{\mathrm{sr}}}\Bigr)^8
\Bigl[\frac{1}{2}\sigma-\sqrt{2\sigma}\Bigl(\frac{R}{R_{\mathrm{sr}}}\Bigr)\Bigr]^{-2}.
\end{equation}
\end{subequations}
These two approximations match at the scale $R_{\mathrm{t}}\approx
\frac{1}{4}\sqrt{\frac{\sigma}{2}}R_{\mathrm{sr}}$ at a value
$K_{\mathrm{t}}\approx 4^{-8}\sigma^2$, and we therefore use the
form~\eqref{eq:K small} for $t<t_{\mathrm{t}}$ and~\eqref{eq:K
long} for $t_{\mathrm{t}}<t<t_{\mathrm{sr}}$.

Since we expect that $\sigma$ is a small parameter, we can now
determine the evolution of the comoving scale factor $R$
throughout the early universe.  In fact, the only new feature is
that for $t<t_{\mathrm{t}}$ we have $K\propto R^8$, which does not
alter the de~Sitter behaviour of the scale factor since we still
have $K\approx 1$.

It is also useful to show that for small $\sigma$ we have
$I_\rho\approx 1$ throughout the history of the universe. To begin
with, we take a time derivative of $I_\rho$ to find (retaining the
scalar field potential for the sake of completeness):
\begin{equation}
{I}_\rho\dot{{I}}_\rho
 = -{\sigma} K Z H [5-9K+(3-K){\sigma} Z]
  -3{\sigma} Z K^2\dot{V}_B.
\end{equation}
Near $R=0$ we have $Z\approx 1/\sigma$ and we easily see that
$I_\rho$ begins near $I_\rho\approx 1$ and decreases until a time
$t_{\mathrm{m}}$ where the right-hand side vanishes (ignoring the
potential), at which point we find
\begin{equation}\label{eq:mid condition}
{\sigma}
Z_{\mathrm{m}}=\frac{9K_{\mathrm{m}}-5}{3-K_{\mathrm{m}}}.
\end{equation}
After this $I_\rho$ will increase again and end up at
$I_\rho\approx 1$ as $Z\rightarrow 0$.  Therefore~\eqref{eq:mid
condition} corresponds to the minimum value that $I_\rho$ will
attain.  By combining~\eqref{eq:mid condition} with~\eqref{eq:Ky}
it is not difficult to see that $K_{\mathrm{m}}\approx
\frac{5}{9}+K_0\sigma^{1/3}$ where $K_0\approx 0.989$.  This gives
$\sigma Z_m\approx 3.64\sigma^{1/3}$ and finally
$I_{\rho}(t_{\mathrm{m}})\approx 1 - 2.02\sigma^{1/3}$.  Clearly
the correction can be neglected for small enough $\sigma$.

The scenario that emerges is the following.  Since we have chosen
to work in comoving coordinates of the matter metric
$\hat{\metric}_{\mu\nu}$, (massless) matter fields propagate with
a velocity $c$ throughout the history of the universe.  For times
$t< t_{\mathrm{t}}$ we have $I_p\approx 4/3$ and so
from~\eqref{eq:biscalar speed} we have the speed of biscalar wave
propagation is $(4/3)\sqrt{K}c$, whereas that of gravitational
modes contained in $\metric_{\mu\nu}$ is $\sqrt{K}c$.  Following
this time there is a period ($t_{\mathrm{t}}<t<t_{\mathrm{pt}}$)
in which the biscalar and gravitational speeds coincide at
$\sqrt{K}c$, and later ($t_{\mathrm{pt}}<t$) all three speeds
coincide and Lorentz symmetry is completely restored.

It is interesting that it is the radiation pressure (rather than
the energy density) that is ultimately determining the transition
at $t_{\mathrm{t}}$.  That is, it is $\kappa B \sqrt{K}p\approx
1/3$ that characterizes the very earliest phase of the universe,
and, while it is responsible for the second splitting of the light
cones, it has very little effect on the de~Sitter behaviour of the
scale factor.  Consequently, although the matter energy
density~\eqref{eq:radiation} evaluated at $t_{\mathrm{t}}$ is
$\rho_{\mathrm{t}}\approx 4^5\rho_B/\sigma$ and is therefore
larger than the Planck density, the effective gravitational
coupling to matter (from~\eqref{eq:Fried eqn}) is $K^{3/2}G$.
Using~\eqref{eq:bar rhoP} we have the effective Planck density for
matter is $\rho_{\mathrm{P}}/K^3$, and evaluated at $t_{\mathrm{t}}$
this becomes $4^{24}\rho_{\mathrm{P}}/\sigma^6$, and we see that
the matter energy density is much less than this.

\section{Conclusions}
\label{sect:conclusion}

We have introduced a bimetric structure in a scalar-tensor
gravitational theory coupled to matter, resulting in different
propagation speeds for gravitational waves and matter waves.  This
can in principle be detected in gravitational wave experiments,
such as LIGO, VIRGO and
LISA~\cite{Flanagan:1998,Abramovici+etal:1992}.

Within this model we have developed an alternative scenario to the
standard inflationary models of cosmology, demonstrating the
existence of a solution to the horizon and flatness problems
without resorting to generic slow-roll scalar field potentials. In
fact, we have shown that the de~Sitter phase in the very early
universe is largely independent of the introduction of a scalar
field potential.  Given ongoing concern about the cosmological
constant problem in cosmology (see, for
example~\cite{Brandenberger:2001}), we feel that this is a very
significant feature of our model.

The model is characterized by a time $t_{\mathrm{pt}}$ before
which the light cone of fields in the gravitational sector is
contracted with respect to matter. (In the presence of matter
fields, there will be an earlier time $t_{\mathrm{t}}$ before
which the light cone of the biscalar field bifurcates from that of
the rest of the gravitational sector.)  Prior to $t_{\mathrm{pt}}$
the universe expands exponentially, driven by the kinetic terms of
the scalar field, and is an effect that is attributable entirely
to the differing light cones of matter and gravity.  As time
passes through $t_{\mathrm{pt}}$ a phase transition occurs, the
light cones collapse back onto themselves, and the universe is
effectively described by a minimally-coupled scalar field.

Matter is coupled to this extended scalar-tensor gravitational
sector in a manner which preserves the Einstein equivalence
principle.  Therefore the matter model lives on a geometry
described by a metric $\hat{\metric}_{\mu\nu}$, and we do not
expect that our model will be in conflict with any direct
equivalence principle tests, nor will it violate charge
conservation in the manner described
in~\cite{Landau+Sisterna+Vucetich:2000}. The strong equivalence
principle is violated, and it is the gravitational reaction
to the presence of matter that is altered.  This is precisely what
provides the apparent equation of state that violates the strong
energy condition, despite the fact that the matter and scalar
field actions are of the standard forms.

Due to the stretching of the perturbative scalar field modes in the early universe in
the inflationary period, we expect that a scale invariant microwave
background spectrum will be predicted, once the initial conditions for the
scalar field and the metric are adopted. This issue will be explored in a later
publication.

\section*{acknowledgements}

This work was partially supported by the Natural Science and Engineering Research
Council of Canada.


\end{document}